%% file: main_PRL_final.tex
\documentclass[aps,prl,twocolumn,superscriptaddress,10pt,article,showpacs,longbibliography]{revtex4-1}
\usepackage[utf8]{inputenc}
\usepackage{amsfonts}
\usepackage{amsmath}
\usepackage{amsthm}
\usepackage{braket}
\usepackage{graphicx}
\usepackage{hyperref}
\usepackage{siunitx}
\hypersetup{
    colorlinks=true,
    linkcolor=blue,
    filecolor=cyan,      
    urlcolor=magenta,
    citecolor=purple,
}
\usepackage[dvipsnames]{xcolor}
\usepackage{amssymb}
\usepackage{dsfont}
\usepackage{makecell}
\usepackage{verbatim}
\usepackage{amsmath,amscd}
\usepackage{multirow}
\usepackage{tabularx}
\usepackage[normalem]{ulem}
\usepackage{bm}
\def \k{{\mathbf k}}
\def \q{{\mathbf q}}

\def \beq{\begin{eqnarray}}
\def \eeq{\end{eqnarray}}
\def \nn{\nonumber \\}

\DeclareMathOperator{\Tr}{Tr}

\definecolor{blue-violet}{rgb}{0.54, 0.17, 0.89}

\newcommand{\supplementarysection}{%
  \setcounter{figure}{0}
  \let\oldthefigure\thefigure
  \renewcommand{\thefigure}{S\oldthefigure}
  \setcounter{section}{0}
  \let\oldthesection\thesection
  \renewcommand{\thesection}{S\oldthesection}
  \setcounter{equation}{0}
  \let\oldtheequation\theequation
  \renewcommand{\theequation}{S\oldtheequation}
  \setcounter{table}{0}
  \let\oldthetable\thetable
  \renewcommand{\thetable}{S\oldthetable}
}

\begin{document}

\title{Electrical control of spin and valley in spin-orbit coupled graphene multilayers}

\author{Taige Wang}
\thanks{These two authors contributed equally}
\affiliation{Department of Physics, University of California, Berkeley, CA 94720, USA}
\affiliation{Material Science Division, Lawrence Berkeley National Laboratory, Berkeley, CA 94720, USA}

\author{Marc Vila}
\thanks{These two authors contributed equally}
\affiliation{Department of Physics, University of California, Berkeley, CA 94720, USA}
\affiliation{Material Science Division, Lawrence Berkeley National Laboratory, Berkeley, CA 94720, USA}

\author{Michael P. Zaletel}
\affiliation{Department of Physics, University of California, Berkeley, CA 94720, USA}
\affiliation{Material Science Division, Lawrence Berkeley National Laboratory, Berkeley, CA 94720, USA}

\author{Shubhayu Chatterjee}
\affiliation{Department of Physics, University of California, Berkeley, CA 94720, USA}
\affiliation{Department of Physics, Carnegie Mellon University, Pittsburgh, PA 15213, USA}

\begin{abstract}
Electrical control of magnetism has been a major techonogical pursuit of the spintronics community, owing to its far-reaching implications for data storage and transmission. 
Here, we propose and analyze a new mechanism for electrical switching of isospin, using chiral-stacked graphene multilayers, such as bernal bilayer graphene or rhombohedral trilayer graphene, encapsulated by transition metal dichalcogenide (TMD) substrates. 
Leveraging the proximity-induced spin-orbit coupling from the TMD, we demonstrate electrical switching of correlation-induced spin and/or valley polarization, by reversing a perpendicular displacement field or the chemical potential. 
We substantiate our proposal with both analytical arguments and self-consistent Hartree-Fock numerics. 
Finally, we illustrate how the relative alignment of the TMDs, together with the top and bottom gate voltages, can be used to selectively switch distinct isospin flavors, putting forward correlated van der Waals heterostructures as a promising platform for spintronics and valleytronics.
\end{abstract}

\maketitle

{\em Introduction --}
Electrical control of magnetic degrees of freedom can pave the way for energy-efficient next-generation solid-state devices \cite{Matsukura2015,Ramesh2021,Ralph2008,Manchon2019, Ramesh2021,Lin2019,wang2013low,Yang2022}. 
However, the physical origin of ferromagnetism lies in the interplay of Coulomb repulsion and Fermi statistics of electrons \cite{auerbachInteractingElectronsQuantum1994,coleman_2015}, thereby making it challenging to directly tune magnetism via applied electric fields. 
Two-dimensional heterostructures, such as those made by graphene, transition metal dichalcogenides (TMDs) and low-dimensional magnets, offer a versatile and highly tunable platform to address this challenge \cite{geim2013van,Ahn2020,andrei2021marvels,Sierra2021,Kurebayashi2022, Huang2018, Jiang2018, Wang2018, Johansen2019}. 
In these materials, a wide range of experimental knobs allows one to significantly enhance electronic interaction strength relative to bandwidth, and thereby engineer a variety of correlated phases including spin and orbital ferromagnets. 
Specifically, electric field switching of orbital magnetism via manipulation of topological edge states has been demonstrated in moir\'e graphene multilayers \cite{DGG_2020_TBGswitch,Zhu2020,Polshyn2020}. 

While moir\'e materials have been in the spotlight for their unprecedented tunability, the moir\'e pattern itself acts as a new source of disorder --- twist angle disorder stemming from local lattice relaxation \cite{Leconte2022}.
As a consequence, the physics of moir\'e systems is often found to vary significantly from device to device \cite{lau2022reproducibility}.
Further, the nature of ordering of spin-degrees of freedom in moir\'e materials has proven difficult to detect and manipulate. 
Very recently, both orbital (valley) and spin ferromagnetism has been observed in few-layered graphene in absence of any moir\'e pattern \cite{Zhou2021_ABCmetals,Zhou_ABCSC,Zhou_BBG, delaBarrera2022,Zhang_BBGSOC,Holleis2023, GuineaReview,han2023correlated,liu2023interactiondriven}.
These remarkable experiments naturally motivate the following questions. 
First, can we exploit the tunability of graphene-based heterostructures to design an electrical switch of magnetism without involving a moir\'e superlattice?
Second, can we selectively switch spin and valley degrees of freedom on demand?

In this paper, we introduce TMD encapsulated chirally stacked graphene as a new platform for electrical control of isospin. 
Specifically, we demonstrate a mechanism to selectively switch valley and/or spin polarization by reversing a perpendicular displacement field, by leveraging both proximity-induced spin orbit coupling (SOC) from TMD substrates, and interaction-induced strong correlations in chiral graphene multilayers.

To substantiate our proposal, we first employ analytical arguments and self-consistent Hartree-Fock (HF) numerics to establish the robust presence of isospin (spin and valley) polarized ferromagnetic phases in the graphene-TMD heterostructure under a displacement field $\vec{D}$.
Next, we show that the degeneracy between different spin-valley polarized phases is split due to the spin-valley locking SOC, and the splitting can be controlled by voltages on top and bottom gates.
Therefore, reversing gate voltages flips valley and/or spin polarization. 
Further, we discuss how the relative alignment of the top and bottom TMD layers and the carrier type (electron vs. hole) may be used to select the precise isospin degrees of freedom that are electrically switched.
\begin{figure}[!htbp]
	\centering
    \includegraphics[width = 0.48\textwidth]{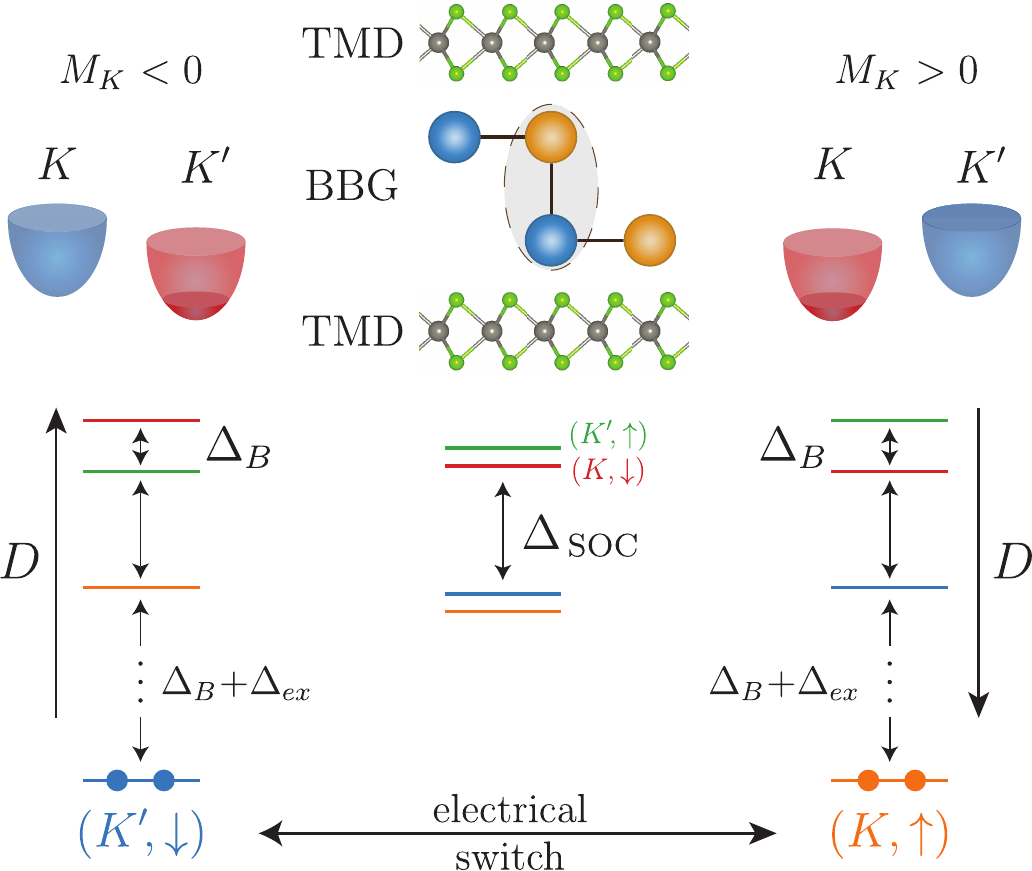}
    \caption{Schematic depiction of the switching mechanism for correlated electrons in the TMD-encapsulated chirally-stacked graphene, with top and bottom TMDs aligned. The SOC-induced splitting, $\Delta_\text{SOC}$, favors a pair of time-reversal related degenerate  $(K, \uparrow)$, $(K^\prime, \downarrow)$ states. Interaction effects polarize carriers spontaneously into one of these two flavors, with a gap $\Delta_{\rm ex} \gg \Delta_{B}$ (gaps not to scale). The Zeeman energy $\Delta_\text{B}$ splits this twofold degeneracy, based on the orbital magnetization $M_\tau$ of electrons in valley $\tau$, and selects a specific flavor polarization. As $M_\tau$ is odd in $\vec{D}$, switching the direction of the displacement field swaps valley polarization. Since valley is locked to spin via SOC, this also leads to a flip of spin polarization.}
    \label{fig:schematics}    
\end{figure}

{\em Switching mechanism --}
We start with a simple physical picture of the switching mechanism, schematically depicted in Fig.~\ref{fig:schematics}.
To this end, we consider chirally stacked graphene (CSG) multilayers, such as Bernal bilayer graphene (BBG) or rhombohedral trilayer graphene (RTG), encapsulated within TMD layers. 
Like monolayer graphene, the low-energy degrees of freedom in CSG comprise of four isospin flavors --- valley ($\tau = K,K^\prime$) and spin ($s = \uparrow,\downarrow$).
Applying a displacement field $\vec{D}$ on CSG enhances the low-energy density of states, and promotes isospin polarization at low carrier density \cite{Zhou2021_ABCmetals,Zhou_ABCSC,Zhou_BBG,delaBarrera2022,Zhang_BBGSOC,Holleis2023, GuineaReview}.
Such a fully isospin polarized phase is spin-valley locked due to the induced SOC (say, $(K,\uparrow)$ or $(K^\prime,\downarrow)$), and carries both spin magnetization $M_s$, and orbital magnetization $M_\tau$, which is constrained by time-reversal to be opposite in the two valleys, i.e., $M_K = - M_{K^\prime}$.

The two-fold degeneracy of spin-valley locked phases may be split by a weak static out-of-plane magnetic field $B$, which couples to both valley and spin degrees of freedom.
Since the orbital Zeeman energy at low carrier density is typically much larger than the spin Zeeman energy, the favored isospin polarized phase corresponds to choosing the valley $\tau$ that has orbital magnetization $M_\tau$ parallel to $B$ (say, $(K',\downarrow)$). 
Therefore, the combination of induced SOC and magnetic field allows one to selectively choose the isospin polarization in CGS.

Having chosen a particular isospin polarization, the central result of our work is that one can switch either valley, spin or both isospin flavors electrically, by flipping the displacement field $\vec{D}$ or the chemical potential $\mu$.
To illustrate this, consider the simplest example where the TMDs are aligned, such that the induced SOC on the top and bottom graphene layers is the same. 
The direction of $\vec{D}$ sets the sign of the Dirac mass, and consequently determines the sign of orbital magnetization which is carrier (electron-hole) and valley dependent \cite{XiaoPRL,NiuRMP}.
On reversing $\vec{D}$ at fixed carrier density, the Dirac mass changes sign, implying that the orbital magnetization of the two valleys are swapped, i.e., $M_K \leftrightarrow M_{K^\prime}$.
To optimize the orbital Zeeman energy, the system prefers to keep the orbital magnetization parallel to $B$  when $\vec{D}$ is reversed. 
This can be achieved by flipping the valley polarization (i.e., $K^\prime \to K$), which in turn leads to a reversal of the spin polarization due to SOC-induced spin-valley locking, (i.e., $(K^\prime, \downarrow) \to (K, \uparrow)$). In this setting, our proposed device acts as a spin and valley switch (Fig.~\ref{fig:schematics}). 
Alternately, one may flip a suitable combination of $\vec{D}$ and $\mu$, and choose the relative alignment of the TMDs appropriately to realize either a spin switch or valley switch, as discussed later and summarized in Table \ref{tab:carrier}.

\begin{figure}[!t]
    \centering
    \includegraphics[width = 0.48\textwidth]{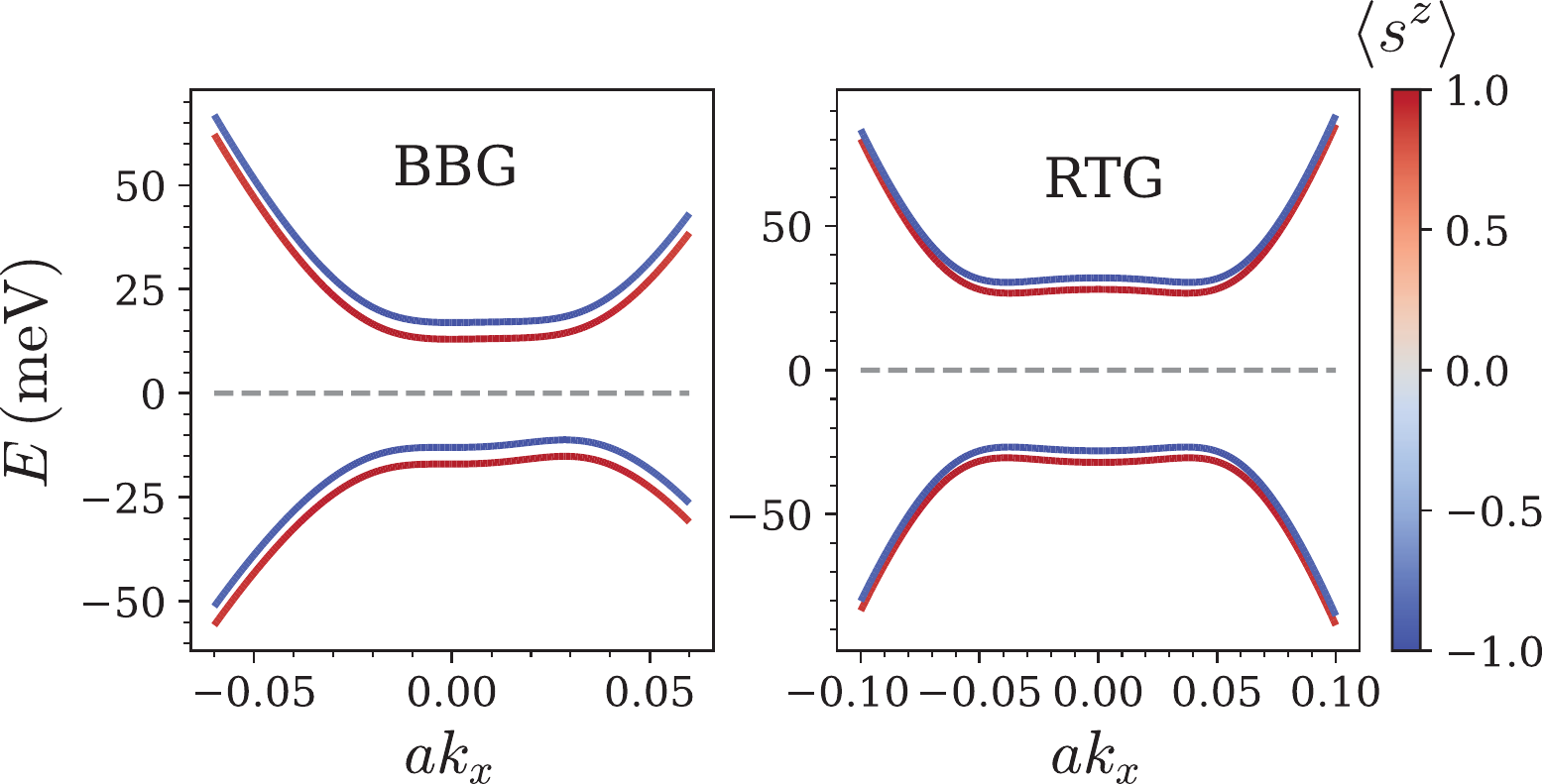}
     \caption{Band structure around $K$ valley of TMD-encapsulated Bernal bilayer graphene (BBG) and rhombohedral trilayer graphene (RTG), with top and bottom TMDs aligned, for typical parameter values: Ising SOC $\lambda_I = \SI{2}{meV}$, Rashba SOC $\lambda_R = \SI{2}{meV}$, and displacement field $u_D = \SI{30}{meV}$. We show the spin polarization $\langle s^z \rangle$ in color. $s^z$ remains a good quantum number near the valence band maxima/conduction band minima at finite $\lambda_R$.
     }
    \label{fig:band}    
\end{figure}

{\em Model --} We begin by considering $N_\ell$-layered CSG in absence of SOC.
While an accurate description of the bandstructure requires $2N_\ell$ bands per valley per spin, in the vicinity of charge neutrality the electron wave-functions in CSG are nearly localized on the $A_1$ and $B_{N_\ell}$ sublattices on the top and bottom layers (Fig.~\ref{fig:schematics}). 
Therefore, an intuitive description may be obtained by employing an approximate 2-band model that describes a single valley/spin flavor in the $\nu = (A_1, B_{N_{\ell}})$ pseudospin basis \cite{McCannReview}. 
The applied displacement field $\vec{D}$ flattens the low-energy dispersion, leading to enhanced interaction effects and isospin flavor polarized phases. 
Multiple such phases, such as a spin-polarized valley-unpolarized half metal, or a spin and valley polarized quarter metal, have already been observed in recent experiments on rhombohedral (ABC) trilayer graphene (RTG) \cite{Zhou2021_ABCmetals,Zhou_ABCSC} and Bernal bilayer graphene (BBG) \cite{Zhou_BBG, delaBarrera2022, Zhang_BBGSOC, Holleis2023,Weitz1,Weitz2}.

\begin{figure*}[htbp]
    \centering
    \includegraphics[width = 0.9\textwidth]{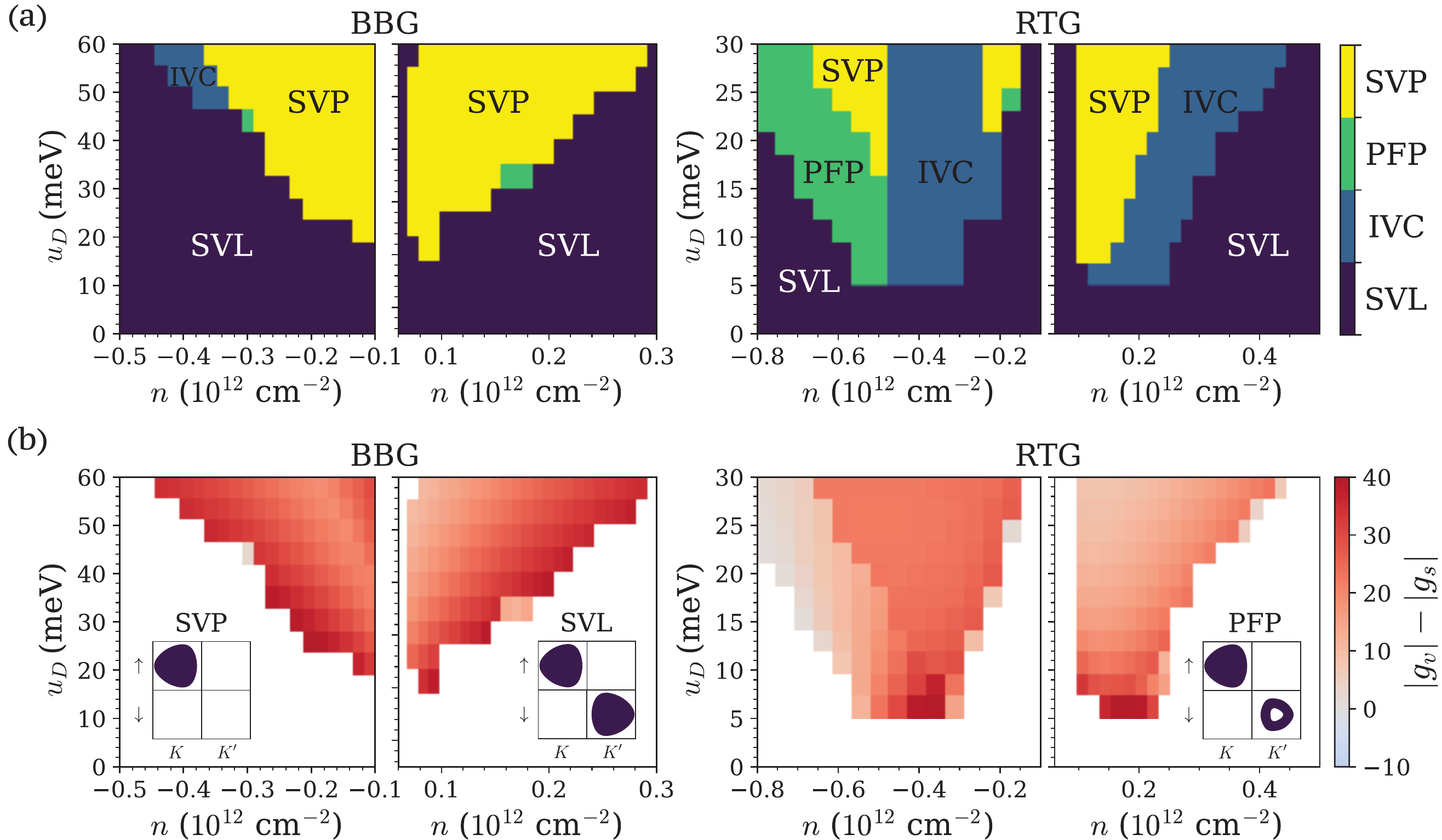}
    \caption{(a) Self-consistent Hartree-Fock phase diagram of TMD-encapsulated BBG and RTG with top and bottom TMDs aligned, for Ising SOC $\lambda_I = \SI{2}{meV}$ and Rashba SOC $\lambda_R = 0$ (small $\lambda_R \neq 0$ does not affect the phase diagram \cite{SM}). The four phases in play are the spin-valley polarized (SVP) phase, the partial flavor polarized (PFP) phase, the inter-valley coherent (IVC) phase, and the symmetry-preserving spin-valley locked (SVL) phase. Panel (b) insets show representative Fermi surface configurations of these phases. (b) The valley $g$-factor $g_v$ of the SVP/PFP phase, when the SVP/PFP phase is favored over the SVL phase. The vast majority of the phase diagram supports $|g_v| > |g_s|$.}
    \label{fig:phase}    
\end{figure*}

When CSG is placed on an insulating TMD substrate, spin-orbit coupling is induced on the graphene layer adjacent to the TMD via virtual tunneling of electrons between graphene and TMD \footnote{The moir\'e superlattice potential is expected to be not important \cite{Gmitra}}. 
The induced SOC on a single layer can be well-approximated by spatially independent Ising and Rashba terms \cite{Gmitra, Gmitra2016, Kochan2017, Garcia2018,Khoo2017,Zollner2022}:
\beq
H_{\rm I-SOC} = \lambda^\ell_I \tau^z s^z \sigma^0, \, H_{\rm R-SOC} = \lambda^\ell_R (\tau^z \sigma^x s^y - \sigma^y s^x) ~~~~~~
\eeq
where $\ell = \rm{t,b}$ denotes the top/bottom graphene layer adjacent to the TMD substrate and $(\tau, \sigma, s)$ denotes Pauli matrices acting on the valley, sublattice and spin degrees of freedom.
Since the Rashba SOC is off-diagonal in the sublattice index and the wave-functions of the low-energy electrons in chiral graphene are nearly sublattice/layer polarized (Fig. \ref{fig:band}(b)), $\lambda_R$ can be neglected.
More explicitly, perturbation theory including both forms of SOC, but neglecting farther neighbor hoppings (e.g., $\gamma_2, \gamma_3, \gamma_4$ in Ref.~\onlinecite{MacDonaldABC}), yields the following effective Hamiltonian in the low-energy $(A_1, B_{N_\ell})$ subspace \cite{KhooZaletel,MacDonaldABC}:
\beq
H_0 &=& \sum_{\tau,s,\k, \nu, \nu^\prime} c^\dagger_{\tau,s,\k, \nu} \left([h_{\tau}(\k)]_{\nu, \nu^\prime} - \mu \, \delta_{\nu \nu^\prime} \right) c_{\tau,s, \k, \nu^\prime} \nn
h_{\tau}(\k) &=& \begin{pmatrix} \lambda_I^{\rm t} \tau^z s^z + u_D & \gamma_1 \left( \frac{v k_-}{\gamma_1} \right)^{N_\ell}  \\ \gamma_1 \left( \frac{v k_+}{\gamma_1} \right)^{N_\ell} & \lambda_I^{\rm b} \tau^z s^z - u_D \end{pmatrix}  
\label{eq:H0}
\eeq
where $k_\pm = \tau^z k_x \pm i k_y$, $v$ is the Dirac velocity for monolayer graphene, $2 u_D \propto N_\ell |\vec{D}|$ is the potential difference between the top and bottom layers due to $\vec{D}$, and $\gamma_1$ is the inter-layer hopping.
We note that only the Ising SOC term $\lambda_I$ appears in $H_0$, and spin $s^z$ remains a good quantum number at low-doping.
This is borne out in a microscopic calculation of the full band structure of BBG/RTG including both Ising and Rashba SOC and realistic hopping parameters \cite{MacDonaldABC}: we find that the low-energy bands in each valley are nearly spin-polarized (Fig.~\ref{fig:band}) but split in opposite directions.
For the anti-aligned device, inversion symmetry constrains $\lambda_I^{\rm t} = - \lambda_I^{\rm b}$, whereas it is reasonable to approximate $\lambda_I^{\rm t} = \lambda_I^{\rm b}$ when the TMDs are aligned  \cite{KhooZaletel}.

{\em Interaction induced flavor polarization --} 
We now determine the interacting phase diagram of TMD encapsulated CSG. 
Just like bare CSG, TMD encapsulated CSG is also expected to exhibit flavor polarized phases at sufficiently large $u_D$, as evidenced by recent experiments on hole-doped BBG on TMD substrates \cite{Zhang_BBGSOC, Holleis2023}. 
To confirm this expectation, we perform self-consistent Hartree-Fock numerics (HF) of TMD encapsulated BBG and RTG at both electron and hole dopings using a realistic $2N_\ell$-band model and screened Coulomb interaction $V_C(\q)$ projected onto the \textit{active} bands \cite{SM}. 
The resulting HF phase diagram is shown in Fig.~\ref{fig:phase}(a) \footnote{The phase diagram remains quantitatively the same when $\lambda_R = \SI{5}{meV}$ as shown in \cite{SM}.}. At small $u_D$, interactions do not induce flavor polarization due to low density of states. 
The corresponding phase is the spin-valley locked (SVL) metal that preserves all symmetries of the non-interacting ground state.
At large $u_D$, we find several symmetry-broken phases.
Of particular interest to us are the valley and spin polarized phases, including the fully spin and valley polarized (SVP) phase (the analogue of the quarter metal phase in RTG without SOC \cite{Zhou2021_ABCmetals}) and the partial flavor polarized (PFP) phase which also has finite valley and spin polarization. All these polarized phases resemble generalized ferromagnets, where different spin/valley species support Fermi surfaces with different volumes, as shown in the inset of Fig.~\ref{fig:phase}(b).
Both the SVP and the PFP phase are observed robustly in the interacting phase diagram. 
A competing correlated valley unpolarized state is the inter-valley coherent (IVC) phase, which is a coherent superposition of two valleys with opposite spins. 

{\em Splitting of isospin degeneracy -} We focus on the ferromagnetic SVP phase, which would be fourfold degenerate in absence of SOC. 
Ising SOC reduces this fourfold degeneracy to twofold by locking spin and valley so that the ground states necessarily have $\tau^z s^z = 1$ (or $-1$, depending on the sign of induced Ising SOC in the CSG layer containing the carriers). 
Our goal is to switch either spin, or valley, or both isospin flavors electrically. 
One may think that the switch could be accomplished by simply reversing the displacement field, so that the carriers are polarized to the opposite layer, where the induced SOC has opposite sign if the top and bottom TMDs are anti-aligned. 
However, since SOC always selects a pair of degenerate spin-valley locked states, we still cannot switch to a given isospin on demand.
Consequently, this still leaves open the central question: how do we selectively control valley versus spin polarization?

The resolution to the above puzzle is to energetically split the twofold degeneracy of the ground state manifold by a perpendicular magnetic field $\vec{B} = B \hat{z}$, which \textit{remains fixed} during the switching process. 
In presence of the magnetic field, the effective Hamiltonian features an additional term
\beq
h_B = -  \mu_B B \left[ g_v(u_D) \frac{\tau^z}{2} - g_s \frac{s^z}{2} \right]
\eeq
where $g_s = 2$ is the spin g-factor, and $g_v(u_D)$ is the valley g-factor that arises from the orbital magnetization $M_\tau$ of electrons in valley $\tau$ \cite{NiuRMP,XiaoPRL}.
For the fully isospin polarized SVP phase, we estimate $g_v$ at carrier density $n$ explicitly within the simplified particle-hole symmetric two-band model described by Eq.~\eqref{eq:H0} \cite{SM}:
\beq \label{eq:M}
 g_v(u_D) = \frac{2 M_{\tau = K}}{|n| \mu_B} = \frac{N_\ell e}{2 \pi |n| \hbar \mu_B} \left[u_D  - |\mu| \text{ sgn}(u_D) \right], ~~~
\eeq
where $|\mu| > |u_D|$. 
Eq.~\eqref{eq:M} shows good agreement with our numerical computation of $g_v$ in the Hartree-Fock ground state (Fig.~\ref{fig:phase}), which implements the full microscopic $2 N_\ell$-band model and accounts for trigonal warping.
Importantly, for typical parameters of filling $n \approx 10^{12}$ cm$^{-2}$ and $u_D \approx 30$ meV, we find $|g_v| \approx 20 \gg g_s$.

We have now gathered all ingredients to determine the complete splitting of isopin polarized phases for any direction of the displacement field.
In presence of $h_B$, flavor polarized states with identical $\tau^z s^z$, say $(K,\uparrow)$ and $(K^\prime,\downarrow)$, get split by $\Delta_B = |g_v - g_s|\mu_B B$.
Crucially, since $|g_v| \gg g_s$, the magnetic field prefers to pin the orbital magnetization.
However, from Eq.~\eqref{eq:M} it follows that reversing $\vec{D}$ at fixed carrier density (i.e., at fixed $\mu$) changes the sign of the orbital magnetization in a fixed valley, as $g_v(-u_D)= -g_v(u_D)$ \cite{PHsym}. 
Consequently, the $h_B$-induced pinning of orbital magnetization requires the carriers to move to the opposite valley.
Therefore, after reversing $\vec{D}$ at a fixed carrier density, the energetically preferred ground state always exhibits a valley polarization opposite to that of the initial state.
It remains to examine the resultant spin polarization, which depends on the relative alignment of the TMD substrates (Table~\ref{tab:carrier}). 

{\em Selective switching of isospin -}
In the device with aligned TMDs, the induced Ising SOC is same on both top and bottom graphene layers, and locks spin and valley identically in both layers.
Therefore, switching $\vec{D}$ and consequently valley polarization automatically implies switching of spin polarization. 
Thus the aligned device can serve as a spin-valley switch.

In the device with anti-aligned TMDs, the induced Ising SOC is opposite in the top and bottom graphene layers, and prefers opposite spin-valley locked pairs.
Consequently, shifting carrier density from top to bottom layer by reversing $\vec{D}$ now implies switching the valley polarization while keeping the spin polarization fixed.
Thus the anti-aligned device can serve as a valley switch.

Finally, to complete the trifecta, we discuss a switching protocol which leaves valley polarization unchanged, but flips spin.
To do so, we consider the device with aligned TMDs, and take advantage of the spin-valley polarized phases being available at both electron doping $n_e$ and hole-doping $n_h = - n_e$ (Fig.~\ref{fig:phase}). 
By manipulating the top and bottom gate voltages, we can electrically tune $\mu \to - \mu$, while keeping $\vec{D}$ fixed, swapping electron-like carriers for hole-like carriers.
Since the valley g-factor depends only on $|\mu|$ (Eq.~\eqref{eq:M}), pinning of the orbital magnetization requires the electronic valley polarization to remain unchanged during the process. 
However, as electrons and holes feel opposite signs of Ising SOC, this protocol leads to a switch of electronic spin polarization.

\begin{table}[]
\renewcommand*{\arraystretch}{1.5}
\newcolumntype{C}{>{\centering\arraybackslash}X}
\centering
\begin{tabularx}{0.75\columnwidth}{CCCC}
\hline \hline
                              &                   & $u_D > 0$         & $u_D < 0$        \\ \hline
\multirow{2}{*}{aligned}      & $\mu < 0$ & $(K', \uparrow)$  &  $(K, \downarrow)$  \\
                              & $\mu > 0$ & $(K', \downarrow)$ &  $(K, \uparrow)$   \\ \hline
\multirow{2}{*}{anti-aligned} & $\mu < 0$ & {$(K', \uparrow)$} &  $(K, \uparrow)$  \\
                              & $\mu > 0$ & $(K', \uparrow)$ &  $(K, \uparrow)$  \\ \hline \hline
\end{tabularx}
\caption{
Electronic isospin polarization for different configurations of the displacement field ($u_D$), chemical potential $\mu$ and relative alignment of the top and bottom TMDs. 
Aligned (anti-aligned) implies that $\lambda_I^{\rm t} = \lambda_I^{\rm b}$ ($\lambda_I^{\rm t} = -\lambda_I^{\rm b}$)
and we have chosen $\lambda_I^{\rm b} < 0$. 
For hole-doping ($\mu < 0$), the active carrier is of the minority species, and has opposite polarization to electrons, e.g., $(K', \downarrow)$ implies a hole-like Fermi surface in $(K, \uparrow)$.}
\label{tab:carrier}
\end{table}

{\em Conclusions and outlook --} We proposed and analyzed an all-electrical route to switch either spin and/or valley polarization in chirally stacked graphene encapsulated in TMD substrates.
Our proposal leverages correlated physics in Van der Waals heterostructures to electrically control spin and orbital degrees of freedom, which is essential for spintronics and orbitronics \cite{Canonico2020,go2021orbitronics,Bhowal2021,kim2022orbital}.
It applies to both electron and hole-doped sides of the phase diagram, since flavor polarized metallic phases, unlike superconductivity, are ubiquitous for both electron and hole doping away from charge neutrality \cite{Zhou2021_ABCmetals}.
Since the phases under consideration are interaction-driven and break discrete symmetries, the magnetic order is stable to thermal fluctuations.
Further, we have utilized the induced SOC from the TMD substrate to achieve a higher degree of tunability compared to moir\'e graphene, as we can choose whether to flip spin and/or valley. 
In addition, chiral graphene multilayers, being very clean and reproducible \cite{Zhou_ABCSC} allows one to bypass difficulties such as twist angle disorder in moir\'e materials.

We conclude with a few comments.
First, the mechanism is compatible with small in-plane Zeeman fields, which may be required to enhance interaction effects \cite{Zhou_BBG}. 
Second, at large $N_\ell$, the interplay of enhanced orbital magnetization which aids the switching protocols, and increased electronic screening which may adversely affect flavor polarization \cite{AregABCA}, is an interesting theoretical problem for future work.  
Finally, we note that our interacting phase diagram of BBG and RTG, for both electron and hole doping in presence of SOC, may be useful to throw light on the variety of correlated phases observed in BBG \cite{Zhang_BBGSOC, Holleis2023} and serve as a guide for future experiments on RTG in presence of a TMD substrate.

\begin{acknowledgments} 
{\em Acknowledgements --} We thank L. Antonio Benítez, Ludwig Holleis, Long Ju and Andrea F. Young for helpful discussions. We thank Yang Zhang for providing the TMD atomic structure figure. 
This work was funded by the U.S. Department of Energy, Office of Science, Office of Basic Energy Sciences, Materials Sciences and Engineering Division under Contract No. DE-AC02-05-CH11231 (Theory of Materials program KC2301). 
M.V. was supported as part of the Center for Novel Pathways to Quantum Coherence in Materials, an Energy Frontier Research Center funded by the US Department of Energy, Office of Science, Basic Energy Sciences.
SC is supported by the ARO through the MURI program (grant number W911NF17-1-0323), and a start-up fund from CMU. 
This research used the Lawrencium computational cluster resource provided by the IT Division at the Lawrence Berkeley National Laboratory (Supported by the Director, Office of Science, Office of Basic Energy Sciences, of the U.S. Department of Energy under Contract No. DE-AC02-05CH11231).

{\em Note added --} Recently, we became aware of preprints on the phase diagram of BBG  \cite{DasSarma_BBGSOC, Zhumagulov2023BBG} and RTG \cite{Koh2023, Zhumagulov2023RTG} in the presence of Ising SOC, which are consistent with our results in Fig.~\ref{fig:phase} and in the Supplemental Materials \cite{SM}. After submission, we become aware of orbital multiferroicity in rhombohedral pentalayer graphene \cite{Han2023}, in which the valley-polarized phase persists at zero displacement fields. In such cases, the valley switch we proposed becomes an electrically switchable magnetic memory, as the valley polarization remains even after turning off the displacement field.

\end{acknowledgments}


\bibliography{PRL}

\newpage

\onecolumngrid

\vspace{0.3cm}

\supplementarysection
 
\input{supp_arXiv.tex}

\vfill

\end{document}

%% file: supp_arXiv.tex
\begin{center}
\Large{\bf Supplemental Material: Electrical control of spin and valley in spin-orbit coupled graphene multilayers}
\end{center}

\section{Band structures and symmetries of chirally stacked graphene with TMD substrates}

\subsection{Band-structure of BBG and RTG without SOC}
In this section, we elaborate the 4-band and 6-band models for BBG $(N_\ell = 2)$ and RTG $(N_\ell = 3)$ respectively. 
The band-structures shown in Fig.~2,  as well as the band-projection of the screened Coulomb interaction, were obtained by using the following microscopic models, and then adding induced SOC from TMD substrates that we discuss later. 

AB-stacked bilayer graphene or Bernal bilayer graphene (BBG) consists of two graphene monolayers. Each unit cell consists of four sites (two sublattice sites $A_i,B_i$ of the honeycomb lattice in each layer $i$), which we index by $(A_1,B_1,A_2,B_2)$. For chiral stacking, the B-site of the first layer $B_1$ is aligned directly on top of the A-site of the second layer $A_2$. Following Ref.~\onlinecite{MacDonaldBBG}, we consider the following 4-band Hamiltonian (per valley, per spin) at low energy to describe the single-particle band structure. 
\beq
h_{4-band} = \begin{pmatrix} u_D & -\gamma_0 f(\k) & \gamma_4 f(\k) & \gamma_3 f^*(\k)\\
-\gamma_0 f^*(\k) & u_D + \Delta & \gamma_1 & \gamma_4 f(\k)\\
\gamma_4 f^*(\k) & \gamma_1 & -u_D + \Delta & -\gamma_0 f(\k)\\
\gamma_3 f(\k) & \gamma_4 f^*(\k) & -\gamma_0 f^*(\k) & -u_D
\end{pmatrix}
\eeq
where $f(\mathbf{k}) \equiv \sum_l e^{-i \mathbf{k} \cdot \delta_l}$, and $\delta_1=a(0,1), \delta_2=a(\sqrt{3} / 2,-1 / 2)$, and $\delta_3=a(-\sqrt{3} / 2,-1 / 2)$ are vectors from $A$-site to $B$-sites. The values of all tight binding parameters of our study are chosen from Ref.~\onlinecite{MacDonaldBBG}.
\begin{equation}
    \left(\gamma_0, \gamma_1, \gamma_3, \gamma_4, \Delta\right)=(2610,361,283,138,15) \; \mathrm{meV}
\end{equation}

ABC-stacked trilayer graphene or rhombohedral trilayer graphene (RTG) consists of three graphene monolayers. Each unit cell consists of six sites (two sublattice sites $A_i,B_i$ of the honeycomb lattice in each layer $i$), which we index by $(A_1,B_1,A_2,B_2,A_3,B_3)$. For chiral stacking, the B-site of the one layer $B_i$ is aligned directly on top of the A-site of the next layer $A_{i+1}$. Following Ref.~\onlinecite{Zhang2010,Zhou2021_ABCmetals}, we consider the following 6-band Hamiltonian (per valley, per spin) at low energy to describe the single-particle band structure. 
\beq \label{eq:6band}
h_{6-band} = \begin{pmatrix} u_D + \delta + \Delta_2 & -\gamma_0 f(\k) & \gamma_4 f(\k) & \gamma_3 f^*(\k) & 0 & \frac{\gamma_2}2\\
-\gamma_0 f^*(\k) & u_D + \Delta_2 & \gamma_1 & \gamma_4 f(\k)& 0 & 0 \\
\gamma_4 f^*(\k) & \gamma_1 & - 2 \Delta_2 & -\gamma_0 f(\k)& \gamma_4 f(\k) & \gamma_3 f^*(\k) \\
\gamma_3 f(\k) & \gamma_4 f^*(\k) & -\gamma_0 f^*(\k) & - 2 \Delta_2 & \gamma_1 & \gamma_4 f(\k)\\
0 & 0 & \gamma_4 f^*(\k) & \gamma_1 & -u_D + \Delta_2 & -\gamma_0 f(\k)\\
\frac{\gamma_2}2 & 0 & \gamma_3 f(\k) & \gamma_4 f^*(\k) & -\gamma_0 f^*(\k) & -u_D +\delta + \Delta_2
\end{pmatrix}
\eeq
The values of all tight binding parameters of our study are chosen from Ref.~\onlinecite{Zhou2021_ABCmetals}.
\begin{equation}
    \left(\gamma_0, \gamma_1, \gamma_2, \gamma_3, \gamma_4, \Delta_2, \delta\right)=(3100,380,-21,290,141,-2.3,-10.5) \; \mathrm{meV}
\end{equation}

\textit{Effective 2-band model -}
In the main text, we often used a simplified 2-band model (per spin per valley) to make analytical progress.
Here, we briefly recall the main features of this model. 
Near charge neutrality, the electron wave-functions in both BBG and RTG (and more generally in chirally-stacked multilayer graphene) are almost localized on the $A_1$ and $B_{N_\ell}$ sublattices on the top and bottom layers (see Fig.~1(a)). 
Hence, useful intuition may be gained from an approximate 2-band model that describes a single valley/spin flavor in the $\nu = (A_1, B_{N_\ell})$ pseudospin basis \cite{KhooZaletel,Zhang2010,McCannReview}.
Neglecting trigonal distortion, the non-interacting Hamiltonian (in absence of SOC) is given by:
\beq
H_0 = \sum_{\tau,s,\k, \nu, \nu^\prime} c^\dagger_{\tau,s,\k, \nu} \left([h_{\tau}(\k)]_{\nu, \nu^\prime} - \mu \, \delta_{\nu \nu^\prime} \right) c_{\tau,s, \k, \nu^\prime}, 
 \text{ where } [h_\tau(\k)]_{\nu, \nu^\prime} = \begin{pmatrix}   u_D & \frac{(v k_-)^{N_\ell}}{\gamma_1^{{N_\ell}-1}}  \\  \frac{(v k_+)^{N_\ell}}{\gamma_1^{{N_\ell}-1}} & - u_D \end{pmatrix}_{\nu \nu^\prime} 
\label{eq:H0Supp}
\eeq
$\mu$ is the chemical potential, $k_\pm = \tau^z k_x \pm i k_y$, and $u_D$ is twice the potential difference between the top and bottom layers due to the applied displacement field $\vec{D}$. 
When the displacement field ($u_D$) is large, then the wave-functions of the carriers in the active band are strongly polarized to the top or bottom layer, depending on the direction of $\vec{D}$, as is evident from the Hamiltonian in Eq.~\eqref{eq:H0Supp}.
This continues to hold in presence of SOC, as shown in Fig.~\ref{fig:layer_polarization}.

The symmetries of the model (when there is no SOC) are discussed, for example, in Ref.~\onlinecite{CWBZ2022}; we recall a couple of key symmetries that are useful for our work.
First, as expected, there is global SU(2)$_s$ spin-rotation symmetry, which gets enhanced to independent spin-rotation symmetries in each valley SU(2)$_K \times$ SU(2)$_{K^\prime}$ if we neglect lattice scale effects. 
Second, in the absence of the displacement field ($u_D = 0$), the Hamiltonian of chirally-stacked graphene is symmetric under inversion $I$.
More explicitly, on the effective two-band model, we have 
$[I,H_0] = 0$ where  $I: \tau^x \nu^x \otimes \k \to -\k$.

\subsection{Effect of TMD substrate on monolayer graphene}
To understand the effect of the transition metal dichalcogenide (TMD) substrate on chirally-stacked graphene, it is useful to first consider its two-fold effect on monolayer graphene. 
First, in spite to the insulating nature of the TMD, virtual tunneling of electrons from graphene into the TMD (and back) can induce significant spin-orbit coupling (SOC). 
Second, due to the mismatch of the lattice scales between graphene and TMD substrate, the graphene electrons feel a moir\'e potential. 
From first principles calculations \cite{Gmitra}, it was argued that the effect of the moir\'e potential is negligible, and the main effect of the TMD on the low-energy band-structure of graphene can be modeled well by a spatially independent SOC and a finite sublattice splitting.
In terms of monolayer graphene degrees of freedom ($\tau,\sigma,s$), these effects take the following form \cite{Gmitra}:
\beq
h_{\rm I-SOC} &=& \lambda_I \tau^z s^z \nn
h_{\rm R-SOC} &=& \lambda_R (\tau^z \sigma^x s^y - \sigma^y s^x) \nn
h_{\rm KM-SOC} &=& \lambda_{KM} \tau^z s^z \sigma^z \nn
h_{\rm SL} &=& u_{SL} \sigma^z
\label{eq:inducedSOC}
\eeq
In Eq.~\eqref{eq:inducedSOC}, $\lambda_I$ is the Ising SOC, $\lambda_R$ is the Rashba SOC, $\lambda_{KM}$ is the Kane-Mele SOC \cite{KM}, and $u_{SL}$ is a sublattice splitting. 
Typically, $\lambda_{I}$, $\lambda_{R}$ and $u_{SL}$ are all meV scale, while $\lambda_{KM}$ is much smaller and usually neglected \cite{Gmitra,Zollner2022}.

We briefly discuss the effect of induced SOC on the symmetries of the monolayer graphene Hamiltonian. 
Barring $u_{SL}$, all terms in Eq.~\eqref{eq:inducedSOC} break the global SU(2)$_s$ spin-rotation symmetry of monolayer graphene. 
In addition, monolayer graphene has a spin-ful $C_{2z}$ symmetry that acts as $i \tau^x s^z \sigma^x \otimes \k \to -\k$. 
$h_{\rm I-SOC}$ and $h_{\rm SL}$ are odd under this $C_{2z}$ rotation, while $h_{\rm R-SOC}$ and $h_{\rm KM-SOC}$ are even under it. 
Thus, rotating the graphene monolayer by 180$^{\circ}$ relative to the TMD substrate will take $\lambda_I \to - \lambda_I$ and $u_{SL} \to -u_{SL}$, while keeping $\lambda_R$ and $\lambda_{KM}$ fixed.
One important takeaway from this observation is that the strength of the induced Ising SOC $\lambda_I$ can be controlled by the relative alignment between the TMD substrate and the adjacent graphene monolayer.

\subsection{Two-sided CSG-TMD heterostructures}

Next, we turn to TMD-CSG-TMD heterostructure with opposite TMD orientation on the two sides of CSG. 
In the most general case, the two-sided heterostructure can have different magnitudes and signs of proximity-induced SOC (and sublattice splitting) on the two sides. 
If the TMD layers are aligned (inversion symmetry $I$ is broken), then we expect that the same sign of SOC will be induced on both top and bottom graphene layers of CSG. 
In contrast, if two TMD layers that encapsulates CSG are anti-aligned, then the structure preserves inversion $I$, and places important constraints the induced SOC.
In this subsection, we derive the allowed SOC terms for the CSG Hamiltonian encapsulated in anti-aligned TMD substrates.

We start by introducing some notation that will be useful for deriving the symmetry constraints on SOC.
Following Ref.~\onlinecite{Tianle2020}, we consider by a $+$ superscript a coupling that has same sign on top and bottom layers, and by a $-$ superscript a coupling that has opposite signs on the top and bottom layers. 
For example, $\lambda^+_{I} = (\lambda^t_{I} + \lambda^b_{I})/2$ denotes Ising SOC with same sign on both top and bottom layers (denoted by t/b respectively, corresponding to $\ell^z = \pm 1$), and $\lambda^-_{I} = (\lambda^t_{I} - \lambda^b_{I})/2$ denotes Ising SOC with opposite sign on the two layers.
Thus, the most general $h_{\rm SOC}$ one can write down is:
\beq
h_{\rm SOC} =\sum_{\ell^z = t/b} (\lambda^+_I + \lambda_I^- \ell^z) \tau^z s^z + (\lambda^+_R + \lambda_R^- \ell^z)(\tau^z \sigma^x s^y - \sigma^y s^x) + (\lambda^+_{KM} + \lambda^-_{KM} \ell^z) \tau^z s^z \sigma^z + (u_{SL}^+ + u_{SL}^- \ell^z) \sigma^z ~~~
\label{eq:twosided}
\eeq

Symmetries that flip layers place constraints on the magnitudes of the terms in Eq.~\eqref{eq:twosided}. 
Therefore, we tabulate the symmetry actions for the point group symmetries of CSG (for $u_D = 0$).
While our explicit symmetry representations in Eq.~\eqref{eq:spinfulSym} are valid only for the two-band effective model, our conclusions are valid more generally. 
\beq
M_x &:& i \tau^x s^x \otimes (k_x, k_y) \to (-k_x, k_y) \nn
C_{3z} &:& e^{i \pi s^z/3} \otimes \k \to C_3[\k] \nn
I &:& \tau^x \nu^x \otimes \k \to -\k 
\label{eq:spinfulSym}
\eeq
Note that CSG has no $C_{2z}$ symmetry, and that $I$ is the only symmetry that flips layer (and sublattice) and therefore flips pseudospin $\nu^z = A_1 \leftrightarrow B_{N_\ell}$  ($C_{2x} = M_x I$, which also flips layers, is not an independent symmetry).

The symmetry actions on $H_{\rm SOC}$ (using $\tau^x \nu^x = \tau^x \ell^x \sigma^x$ for inversion $I$, as inversion flips valley, sublattice and layer) are presented in Table \ref{tab:SymSOC}.
Specifically, for the inversion symmetric (anti-aligned) CSG-TMD heterostructure, only the following couplings are allowed in the Hamiltonian:
\beq
h_{\rm SOC, anti-aligned} = \sum_{\ell^z = t/b} \lambda_I^- \ell^z \tau^z s^z + \lambda_R^- \ell^z (\tau^z \sigma^x s^y - \sigma^y s^x) + \lambda^+_{KM}  \tau^z s^z \sigma^z + u_{SL}^- \ell^z \sigma^z
\eeq

\begin{table}[t!]
\centering    \renewcommand{\arraystretch}{1.4}
    \begin{tabular}{>{\centering\arraybackslash}p{1cm}>{\centering\arraybackslash}p{1cm}>{\centering\arraybackslash}p{1cm}>{\centering\arraybackslash}p{1cm}>{\centering\arraybackslash}p{1cm}}
    \hline \hline
    & $\mathcal{T}$ & $M_x$ & $C_{3z}$ & $I/C_{2x}$  \\
    \hline 
    $\lambda_I^+$   & \checkmark & \checkmark & \checkmark  & $\times$ \\
    $\lambda_R^+$   & \checkmark & \checkmark & $\times$ & $\times$ \\
    $\lambda_{KM}^+$ & \checkmark & \checkmark & \checkmark & \checkmark \\
    $u^+$ & \checkmark & \checkmark & \checkmark & $\times$ \\
    $\lambda_I^-$     & \checkmark & \checkmark & \checkmark & \checkmark \\  
    $\lambda_R^-$   & \checkmark & \checkmark & $\times$ & \checkmark \\
    $\lambda_{KM}^-$ & \checkmark & \checkmark & \checkmark & $\times$ \\
    $u^-$ & \checkmark & \checkmark & \checkmark & \checkmark \\
    \hline \hline
    \end{tabular}
    \caption{Symmetry signatures of the various terms induced by TMDs on CSG. $\checkmark$ indicates that the term preserves the symmetry. For inversion $I$ (and $C_{2x}$), $\times$ indicates that the term is odd under the symmetry transformation (i.e., goes to negative of itself), while for $C_{3z}^\prime$ it merely indicates that the symmetry is broken.}
    \label{tab:SymSOC}
\end{table}

Note that the allowed sublattice potential $u^-_{SL} \ell^z \sigma^z$ takes the same value on the active sublattices $A_1/B_{N_\ell}$ for the low-energy bands, and henceforth we will assume that it can be absorbed into the chemical potential $\mu$.
We will also assume that the displacement field $\vec{D}$ does not affect the atomic orbitals in the individual layers, so its effect can be captured by simply adding the $u_D \ell^z$ term to the Hamiltonian without modifying the induced SOC from the substrate. 

For the device with aligned TMDs, inversion symmetry is broken and therefore no such strict symmetry constraints can be placed. 
Nevertheless, since SOC is induced only on the graphene layer directly adjacent to the TMD, we may expect the induced SOC to predominantly have the same sign in both top and bottom layers.
Therefore, we may approximate the Hamiltonian for the aligned device as:
\beq
h_{\rm SOC, aligned}  \approx \sum_{\ell^z = t/b} \lambda_I^+ \tau^z s^z + \lambda_R^+  (\tau^z \sigma^x s^y - \sigma^y s^x) + (\lambda^+_{KM} + \lambda^-_{KM} \ell^z) \tau^z s^z \sigma^z + ( u_{SL}^+ + u_{SL}^- \ell^z) \sigma^z
\eeq
Once again, we can set $\ell^z \sigma^z \approx 1$ for the low-energy valence and conduction bands, implying that $u_{SL}^-$ can be absorbed into the chemical potential, and $\lambda^{-}_{KM}$ just adds to the Ising SOC. 
Further, the effect of $u^{+}_{SL}$ can be absorbed in the displacement field induced potential $u_D$.
Hence, only the Ising and Rashba SOC terms need to be considered for the aligned device.

Putting all this information together, we arrive at the Hamiltonian used to numerically determine the band-structure for BBG (or RTG) encapsulated in TMDs:
\beq \label{eq:full}
h_{BBG/RTG} = h_{4-band/6-band} + h_{\rm SOC, aligned/anti-aligned}
\eeq

\begin{figure}[!t]
    \centering
    \includegraphics[width = 0.75\textwidth]{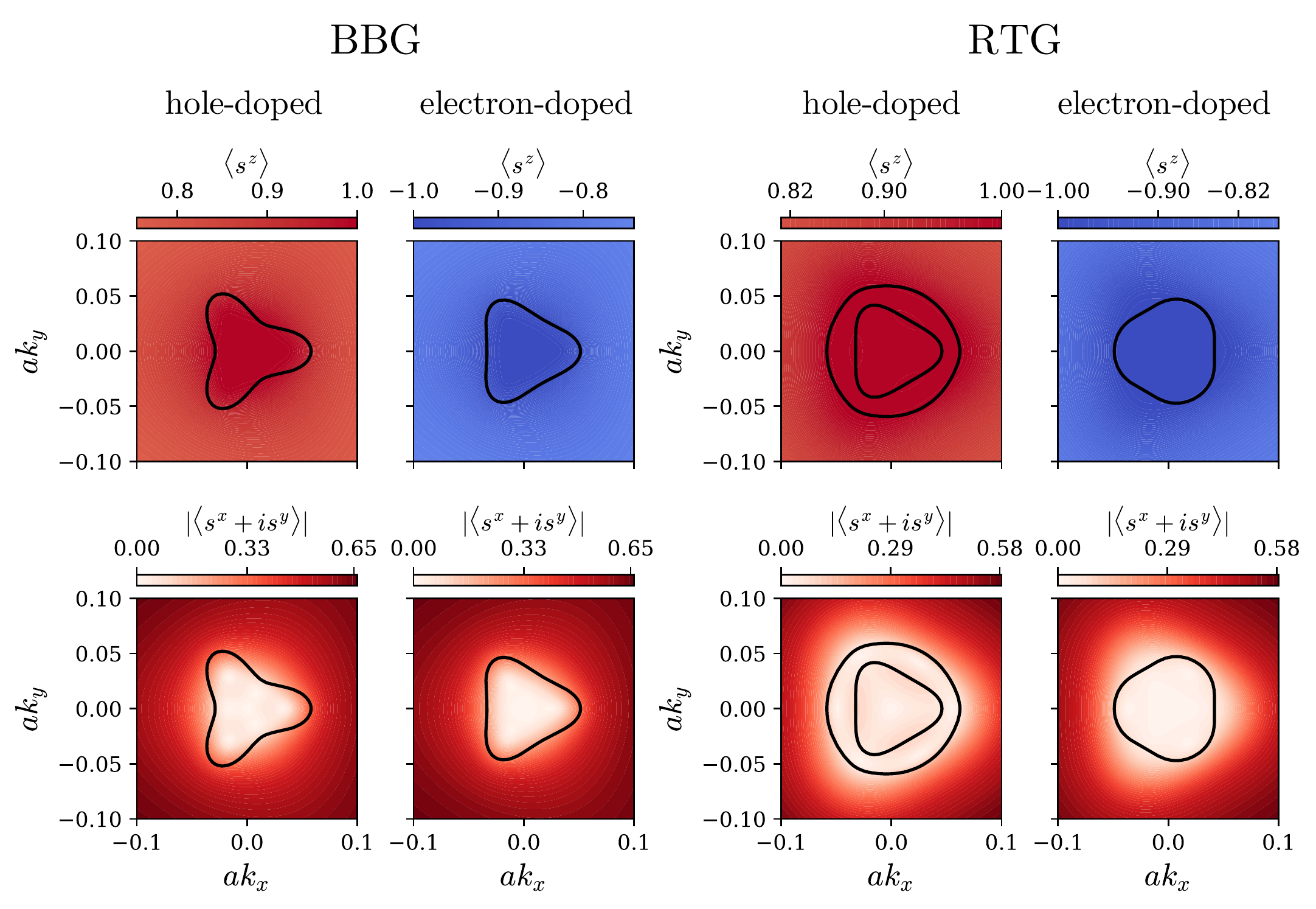}
    \caption{Expectation values $\langle s^z \rangle$ and $\langle |s^x + i s^y| \rangle$ in the K valley \textit{active band} (conduction band for electron-doped and valence band for hole-doped.) of encapsulated BBG and RTG near charge neutrality, using typical values $u_D = 30$ meV, $\lambda_I^{\rm t} = \lambda_I^{\rm b} = 2$ meV and $\lambda_R^{\rm t} = \lambda_R^{\rm b} = 2$ meV (TMD aligned device). 
    The black contour shows the Fermi surface for a fully spin-valley polarized (SVP) phase at carrier denisty $|n| = 0.2 \times 10^{12}$ cm$^{-2}$. 
    We note that the effect of the Rashba term is important only for large carrier densities (the same holds true for the TMD anti-aligned device with $\lambda_I^{\rm t} =- \lambda_I^{\rm b}$).
    This justifies our assumption of approximate $s^z$ conservation to find the interacting phase diagram for low carrier densities that are of interest to us.}
    \label{fig:spin_polarization}    
\end{figure}

\begin{figure}[!t]
    \centering
    \includegraphics[width = 0.5\textwidth]{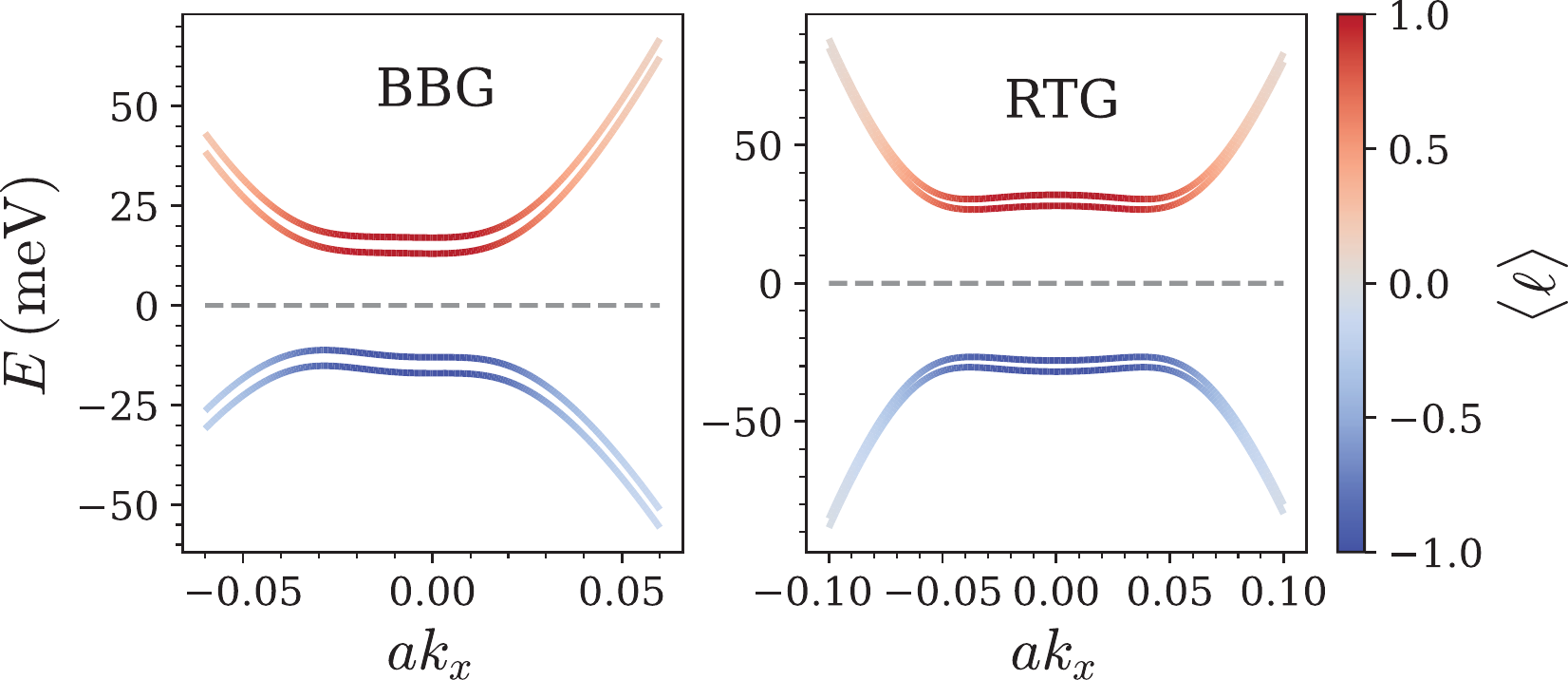}
    \caption{Layer polarization $\langle \ell \rangle = \langle n^{\rm t}_\k - n^{\rm b}_\k \rangle$ overlaid on the band structure around $K$ valley of TMD-encapsulated BBG and RTG, with top and bottom TMDs aligned. The parameters are identical to those used in Fig.~\ref{fig:spin_polarization}. Note that the states near the conduction band minima or valence band maxima are almost entirely layer polarized.}
    \label{fig:layer_polarization}   
\end{figure}

\textit{Effective SOC in the 2-band model -}
As discussed in the main text, the Rashba term, being off-diagonal in the sublattice index $\sigma$, does not contribute significantly to the electronic band structure in the vicinity of charge neutrality. 
In fact, Ref.~\onlinecite{KhooZaletel} showed that on band-projection, to leading order in perturbation theory, its contribution is proportional to $\lambda_R^+ \propto \lambda_R^t + \lambda_R^b$, which is constrained to be zero in inversion-symmetric (TMD anti-aligned) device.
Although it appears at linear order for the device with aligned TMDs, its contribution is proportional to $k^{N_\ell -1}$ (where $\k$ is momentum measured relative to K/K$^\prime$) to leading order in perturbation theory, and therefore quite small.
In Fig.~\ref{fig:spin_polarization}, we plot $\langle s^z \rangle$ in the conduction and valence band for both BBG and RTG using the full micropscopic band structure, and show that $s^z$ is a good quantum number for low carrier density.
Thus, within our two-band effective model we need to consider only the induced Ising SOC, since the sublattice splitting can be absorbed within the chemical potential $\mu$ and the Kane-Mele SOC $\lambda_{KM}$ is negligible. 
These considerations lead to the following simplified two-band Hamiltonian:
\beq
H_0 = \sum_{\tau,s,\k} c^\dagger_{\tau,s,\k, \nu} \left([h_{\tau}(\k)]_{\nu, \nu^\prime} - \mu \, \delta_{\nu \nu^\prime} \right) c_{\tau,s, \k, \nu^\prime}, \text{ where }
h_{\tau}(\k) = \begin{pmatrix} \lambda^{\rm t}_I \tau^z s^z + u_D & \frac{(v k_-)^{N_\ell}}{\gamma_1^{N_\ell-1}}  \\  \frac{(v k_+)^{N_\ell}}{\gamma_1^{N_\ell-1}} & \lambda_I^{\rm b} \tau^z s^z - u_D \end{pmatrix}
\eeq
which is Eq.~(2) in the main text.

\section{Interaction Hamiltonian and self-consistent Hartree-Fock numerics}

\subsection{Hartree-Fock setup}

In this section, we provide details for the Hartree-Fock numerics, which were used to provide interacting phase diagrams of CSG-TMD heterostructures in Fig.~3(a) in the main text. 
The interacting Hamiltonian $H_{\rm C}$ is given by:
\beq
H_{\rm C} = \frac{1}{2A} \sum_{\q} V_C(\q) :\rho(\q) \rho(-\q):
\eeq
where $A$ is the sample area, $V_C(\q) = e^2 \tanh{(q d)}/(2 \epsilon \epsilon_0 q)$ is the dual gate-screened Coulomb interaction with sample-gate distance $d$, and 
\begin{equation}
    \rho(\q) = \sum_{\k,\tau,s,\sigma} c^\dagger_{\tau,s,\k,\sigma} c_{\tau,s,\k+\q,\sigma}
\end{equation}
is the Fourier transform of the electron density operator (neglecting inter-valley scattering which is suppressed for long-range interactions). 
Here, we restrict $|\mathbf{k}|$ and $|\mathbf{q}|$ to be much smaller than the inverse lattice spacing $a^{-1}$.

To study the ground state of this interacting Hamiltonian, we perform a self-consistent Hartree-Fock calculation based on the full $2N_\ell$ band model with both spin and valley flavors, and SOC on the top and bottom layer in Eq.~\eqref{eq:full}. 
For this purpose, we first transform to the band-basis (labeled by electrons $\psi_{n,\tau,s,\k}$), and project the the density operator $\rho(\q)$ in the interaction Hamiltonian to the \textit{active} bands (two conduction or valence bands per valley).
The Hartree-Fock calculation tends to overestimate the exchange energy-gain, as it neglects the screening of interactions by itinerant electrons. To account for this additional screening, we employ the random phase approximation (RPA) for itinerant fermions \cite{coleman_2015},
\beq
H_C \to \frac{1}{2A} \sum_\q V_{\rm RPA}(\q) :\rho(\q) \rho(-\q): \; , \text{ where }  V_{\rm RPA}(\q) = \frac{V_C(\q)}{1 + \chi_{\rho \rho}(\q) V_C(\q)}
\label{eq:HRPASM}
\eeq
At low energy and momentum, we may neglect both frequency and momentum dependence of the screening, and set $\chi_{\rho \rho}(\q) \approx \chi_0$.

In our Hartree-Fock calculation, we solve the self-consistent equations for Slater determinant states characterized by the one-electron covariance matrix $P_{\tau, \tau^{\prime}; s, s^\prime}^{n n^{\prime}}(\mathbf{k})=\left\langle\psi_{n, \tau, s, \mathbf{k}}^{\dagger} \psi_{n', \tau^{\prime}, s^\prime, \mathbf{k}}\right\rangle$, in the band basis labeled by $n$, following the approach outlined in Ref.~\onlinecite{NickPRX}. 
We only consider kinetic energy and intra-valley Coulomb scattering in the Hartree-Fock calculation. 
To numerically solve the self-consistency equations, we employ both the `ODA' and `EDIIS' algorithms \cite{ODA,EDIIS}. The numerical results shown in the paper are obtained with $\epsilon = 4.4$, gate distance $d=\SI{50}{nm}$, using the projected valance band per spin per valley on a $71 \times 71$ momentum grid, with UV momentum cutoff $0.085 a^{-1}$. we use the non-interacting density of states $\chi_0$ with twofold isospin degeneracy, i.e., $\chi_0 = \SI{0.02}{eV^{-1}}$ for BBG and $\chi_0 = \SI{0.08}{eV^{-1}}$ for RTG, which qualitatively reproduces the experimental phase diagram in the absence of SOC \cite{CWBZ2022}.

\begin{figure}[!t]
    \centering
    \includegraphics[width = 0.45\textwidth]{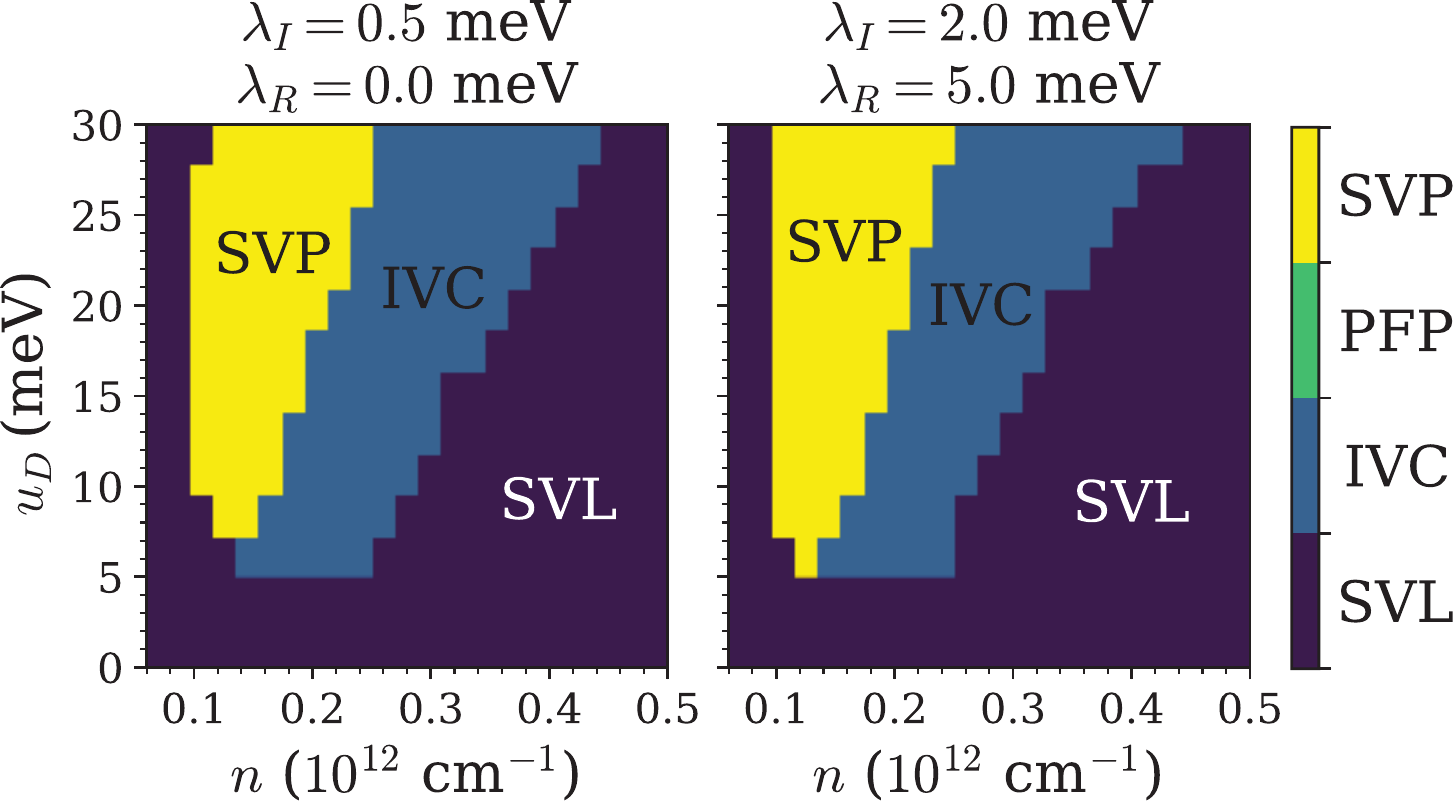}
    \caption{Self-consistent Hartree-Fock phase diagram of TMD encapsulated RTG with different SOC, in the case where the top and bottom TMDs are aligned.}
    \label{fig:phase_diagram_SOC}    
\end{figure}

\subsection{Interacting phase diagram at different SOC}

In this subsection, we study the effects of varying the Ising and Rashba SOC on the phase diagram in Fig.~3(a) in the main text. 
Specifically, we show in Fig.~\ref{fig:phase_diagram_SOC} that the main phases of interest remain almost unchanged, thereby confirming the robustness of our proposal.

\begin{figure}[!t]
    \centering
    \includegraphics[width = 0.82\textwidth]{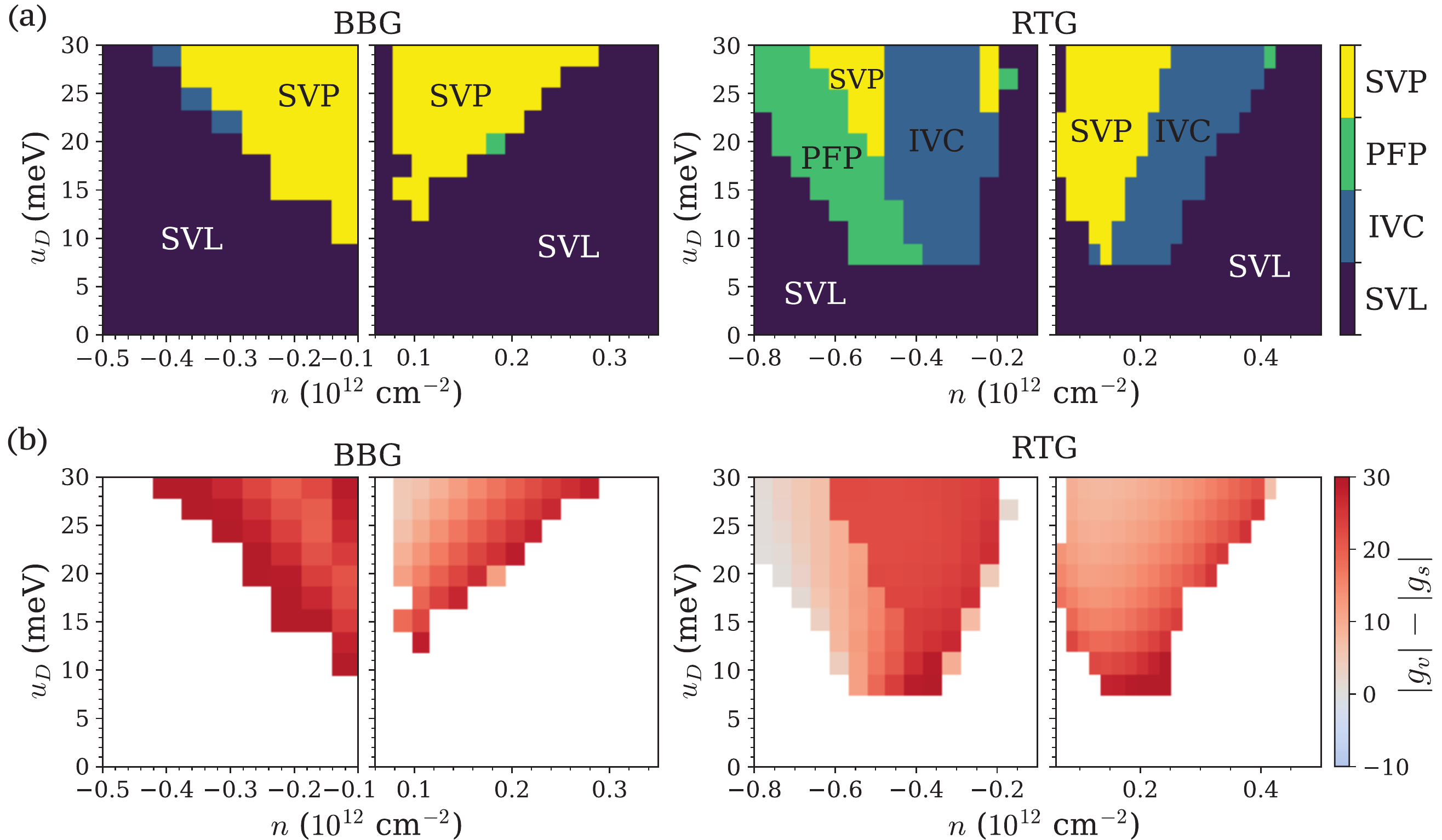}
    \caption{Self-consistent Hartree-Fock phase diagram of TMD encapsulated RTG where the top and bottom TMDs are antialigned ($\lambda_I^{\rm t} = - \lambda_I^{b}$). We chose Ising SOC $|\lambda_I^{\rm t}| = |\lambda_I^{\rm b}| = \SI{2}{meV}$ and Rashba SOC $\lambda_R = 0$ (since its value minimally affects the phase diagram).}
    \label{fig:phase_diagram_anti}    
\end{figure}

The reported Ising SOC from first principle calculations and constraints from experiment phase diagrams vary from $\SI{0}{meV}$ to $\SI{2}{meV}$ \cite{Zollner2022,Zhang_BBGSOC,IsingSOC1,IsingSOC2}. In the main text, we consider a relatively large Ising SOC $\lambda_I = \SI{2}{meV}$. Now we consider a relatively small Ising SOC $\lambda_I = \SI{0.5}{meV}$. Taking electron-doped RTG as an example, we show its Hartree-Fock phase diagram in Fig.~\ref{fig:phase_diagram_SOC}(a). Compared to the phase diagram with $\lambda_I = \SI{2}{meV}$ in Fig.~3(a) in the main text, a smaller $\lambda_I$ favors an IVC phase at large doping, as expected \cite{CWBZ2022}. 
The SVP phase which plays the central role in our proposal is nevertheless untouched.

We also explore the effect of Rashba SOC beyond perturbation theory. The reported Rashba SOC varies in a large range from $\SI{1}{meV}$ to $\SI{15}{meV}$ \cite{Zollner2022,Zhang_BBGSOC,RashbaSOC1,RashbaSOC2,RashbaSOC3}. 
We consider an intermediate value $\lambda_R = \SI{5}{meV}$, keeping $\lambda_I = \SI{2}{meV}$. 
The electron-doped RTG phase diagram in Fig.~\ref{fig:phase_diagram_SOC}(b) closely mimics the phase diagram with $\lambda_R = 0$ in Fig.~\ref{fig:phase_diagram_SOC}(a), justifying our perturbative approach in the main text.

Finally, we consider the device with anti-aligned TMDs on the top and bottom layer, which was used in our protocol for valley switching at fixed carrier density (equivalently, fixed $\mu$). 
Via a self-consistent Hartree-Fock calculation with opposite signs of Ising SOC in the top and bottom layers, i.e.,  $\lambda_I^{\rm t} = - \lambda_I^{\rm b}$, we demonstrate in Fig.~\ref{fig:phase_diagram_anti}(a) that the interacting phase diagram is quite similar to the case when the TMDs are aligned (Fig.~3(a) in the main text).
Further, by numerically evaluating the orbital magnetization in isopsin polarized phases for the anti-aligned device in Fig.~\ref{fig:phase_diagram_anti}(b), we confirm that $|g_v| \gg g_s$ holds irrespective of alignment or anti-alignment of the encapsulating TMDs. 
Taken together, our results justify the use of the TMD anti-aligned device for valley switching, as discussed in the main text. 

\section{Berry curvature and valley $g$-factor}

\subsection{Analytical evaluation}
In this section, we analytically evaluate the valley g-factor $g_v$ for the effective two-band model, using the simplified Hamiltonian in Eq.~(2), which neglects trigonal distortions and particle-hole symmetry-breaking.
Recall that for a fully valley polarized phase, the valley g-factor for the active band is given in terms of the magnetization (density) $M_\tau = \tau^z M$ in valley $\tau = K/K^\prime$ as
\beq 
\frac{\langle h_{\rm valley-Zeeman} \rangle}{N} = - g_v \mu_B B \frac{\tau^z}{2} = -\tau^z  \left( \frac{M B A}{N} \right) \implies g_v = \frac{2 M A}{\mu_B N} =  \frac{2 M}{\mu_B \, n} 
\label{eq:gvFP}
\eeq
where $A$ is the area and $N$ is the total number of carriers, such that $n = N/A$ is the carrier-density. 
(our normalization condition is such that $g_v$ is dimensionless and can be compared to $g_s$).
More generally, for partially valley-polarized (or valley-unpolarized) phases, we may define the valley g-factor by the net orbital coupling to the external magnetic field.
\beq
g_v = \frac{2(\mathcal{M}_K - \mathcal{M}_{K^\prime})A}{\mu_B N} = \frac{2(\mathcal{M}_K - \mathcal{M}_{K^\prime})}{\mu_B \, n}
\label{eq:Mtog}
\eeq
where we have denoted the orbital magnetic moment of the partially filled $\tau$ valley by $\mathcal{M}_\tau$ to differentiate it from the magnetic moment $M_\tau$ in the fully valley polarized state. 
Note that Eq.~\eqref{eq:Mtog} reduces to Eq.~\eqref{eq:gvFP} for the fully valley-polarized case, which has the largest magnitude of $g_v$.
In contrast, if the occupancies of the two-valleys are equal, so are the magnetization densities $\mathcal{M}_K$ and $\mathcal{M}_{K^\prime}$, indicating that $g_v = 0$.
For partially valley-polarized phases, $g_v$ assumes an intermediate value.

We now consider the fully spin and valley polarized phase, and compute $M$ for the K valley (i.e., $\tau^z = 1$); for the $K^\prime$ valley ($\tau^z = -1$) the magnetization has the same magnitude but opposite sign due to spinless time-reversal (when $\lambda_I = 0$). 
The generalization to $\lambda_I \neq 0$ is straightforward, and does not affect the final result significantly. 

For a Bloch Hamiltonian given by $h(\k)$ with eigenstates $\varepsilon_n(\k)$, the magnetization of the n$^{th}$ band is given by \cite{thonhauser2011review}:
\beq \label{eq:Msupp}
M_n =\frac{e}{2\hbar} \left[ \textrm{Im} \int \frac{d^2k}{(2\pi)^2}  \, n_F(\varepsilon_n(\k))  \, \varepsilon_{\mu \nu} \bra{\partial_{k_\mu} u_{n,\k}}  (h(\k) + \varepsilon_n(\k) - 2 \mu) \ket{\partial_{k_\nu} u_{n,\k}}  \right]
\eeq
where $n_F(\varepsilon) = (e^{\beta (\varepsilon - \mu)} + 1)^{-1}$ is the Fermi function.
We can further use the following identity to simplify our calculations:
\beq
\partial_{\k} [(h(\k) - \varepsilon_n(\k))\ket{u_{n,\k}}] = \partial_{\k}[0] = 0 \implies  \ket{\partial_\k u_{n,\k}} = - \sum_{m \neq n} \frac{\ket{u_{n,\k}} \bra{u_{n,\k}} \partial_\k h(\k) \ket{u_{m,\k}}}{\varepsilon_m(\k) - \varepsilon_n(\k)}
\label{eq:BD}
\eeq
Using Eq.~\eqref{eq:BD}, we can express $M_n$ as follows: 
\beq
M_n = \frac{e}{\hbar} \textrm{Im}\left[ \int \frac{d^2k}{(2\pi)^2} \, n_F(\varepsilon_n(\k)) \sum_{m \neq n} \left[ \frac{ \bra{u_{n,\k}} \partial_{k_x} h(\k) \ket{u_{m,\k}} \bra{u_{m,\k}} \partial_{k_y} h(\k) \ket{u_{n,\k}} }{ (\varepsilon_m(\k) - \varepsilon_n(\k))^2 } \right] (\varepsilon_m(\k) + \varepsilon_n(\k) - 2 \mu) \right]
\eeq
where the sum on $m$ runs over all other bands $m \neq n$. 
In our approximate two-band model described by Eq.~\eqref{eq:H0Supp} (we set $\lambda_I = 0$ for simplicity), $M_n$ takes a particularly simple form due to the presence of particle-hole symmetry.
Choosing $n = +$ (conduction band), we can only choose $m = -$ (valence band), implying that $\varepsilon_m(\k) + \varepsilon_n(\k) = \varepsilon_+(\k) + \varepsilon_-(\k) = 0$.
Therefore, we have the following formula for magnetization of the conduction band \cite{XiaoPRL,NiuRMP}:
\beq
M_{+,\tau} = \frac{e \mu}{\hbar} \int \frac{d^2k}{(2\pi)^2} \, n_F(\varepsilon_+(\k)) \Omega_{+,\tau}(\k) = \frac{e \mu \tau^z}{\hbar} \int \frac{d^2k}{(2\pi)^2} \, n_F(\varepsilon_+(\k)) \Omega_{+,K}(\k)
\eeq
where $\Omega_{+,\tau}(\k)$ is the Berry curvature of the conduction band in valley $\tau$, given by \cite{NiuRMP}
\beq
\Omega_{+,\tau}(\k) &=&  i \varepsilon_{\mu \nu} \langle{\partial_{k_\mu} u_{+,\tau,\k}}|{\partial_{k_\nu}  u_{+,\tau,\k}} \rangle = - 2 \, \textrm{Im}[\langle \partial_{k_\mu} u_{+,\tau,\k} |\partial_{k_\nu}  u_{+,\tau,\k} \rangle] \nn 
&=& - 2 \, \textrm{Im} \left[ \frac{ \bra{u_{+,\tau,\k}} \partial_{k_x} h(\k) \ket{u_{-,\tau,\k}} \bra{u_{-,\tau,\k}} \partial_{k_y} h(\k) \ket{u_{+,\tau,\k}} }{ (\varepsilon_+(\k) - \varepsilon_-(\k))^2 } \right] ~~~~~~
\label{eq:BC}
\eeq
For the two band-model in Eq.~(2), we may explicitly evaluate the magnetization density for a given filling of the conduction band.
If we re-write $h(\k) \propto \vec{\nu} \cdot \hat{n}(\theta_\k, \phi_\k)$, where $\hat{n}$ is a unit-vector with angular coordinates $(\theta_\k, \phi_\k)$ on the Bloch sphere, the conduction (+) and valence (-) band energies and wave-functions are given by (for $\lambda_I = 0$):
\beq
\varepsilon_{\pm}(\k) = \pm \sqrt{u_D^2 + \left( \frac{(vk)^{N_\ell}}{\gamma_1^{N_\ell-1}} \right)^2}, ~~~ && \ket{u_{+, \k}} =  \begin{pmatrix} \cos\left(\frac{\theta_\k}{2}\right) \\ e^{i \phi_\k} \sin\left(\frac{\theta_\k}{2}\right) \end{pmatrix}, ~~~ \ket{u_{-, \k}} =  \begin{pmatrix} \sin\left(\frac{\theta_\k}{2}\right) \\ -e^{i \phi_\k} \cos\left(\frac{\theta_\k}{2}\right) \end{pmatrix} \nn
\text{ where } \cos(\theta_\k) &=& \frac{u_D}{\sqrt{u_D^2 + \left( \frac{(vk)^{N_\ell}}{\gamma_1^{N_\ell-1}} \right)^2}}, ~ \phi_\k = N_\ell \, \arctan \left( \frac{k_y}{k_x} \right) \text{ and } k = |\k|
\label{eq:2BES}
\eeq
We plug the eigen-energies and eigenstates from Eq.~\eqref{eq:2BES} into Eq.~\eqref{eq:BC} to find the Berry curvature for the conduction band in the $K$ valley.
\beq
\Omega_{+,K}(\k) = -\frac{ N_\ell^2}{2} \left(  \frac{v^{2N_\ell} k^{2(N_\ell-1)}}{\gamma_1^{2(N_\ell-1)}} \right) \frac{u_D}{\left[ u_D^2 + \left( \frac{(vk)^{N_\ell}}{\gamma_1^{N_\ell-1}} \right)^2 \right]^{3/2}} \label{eq:berry}
\eeq
We note that the Berry-curvature in the opposite valley $K^\prime$ can via $\Omega_{+,K^\prime}(\k) = - \Omega_{+,K}(-\k)$ via time-reversal symmetry. 

Now, we consider the scenario where all carriers in the conduction band reside in one spin and valley flavor.
For our simplified Hamiltonian, tuning the chemical potential $\mu = \mu_c$ to lie in the conduction band leads to a circular Fermi surface with Fermi momentum $k_F$ given by:
\beq
\mu_c = \varepsilon_+(k_F) = \sqrt{u_D^2 + \left( \frac{(vk_F)^{N_\ell}}{\gamma_1^{N_\ell-1}} \right)^2  } =  \sqrt{u_D^2 +  \frac{(4 \pi v^2 n)^{N_\ell}}{\gamma_1^{2(N_\ell-1)}}  }, ~~  \text{ where }  n = \frac{\pi k_F^2}{(2 \pi)^2} = \frac{k_F^2}{4\pi} \text{ is the carrier density} ~~~
\eeq
Consequently, the orbital magnetization in a SVP phase for electron-doping is given by:
\beq
M_{+,\tau} = \frac{e \mu_c}{\hbar} \int_{k \leq k_F} \frac{d^2k}{(2\pi)^2} \, \Omega_{+,\tau}(\k) = -\frac{N_\ell e \tau^z u_D}{4 \pi \hbar} \left(\frac{\mu_c}{|u_D|} - 1\right) = \frac{N_\ell e \tau^z }{4 \pi \hbar}[u_D - \mu \, \text{sgn}(u_D)]
\eeq
For hole doping, an analogous calculation may be used to calculate the magnetization for the unoccupied states. 
In this case, the chemical potential $\mu_v$ lies in the valence band, and is given by
\beq
\mu_v = \varepsilon_-(k_F) = -\sqrt{u_D^2 + \left( \frac{(vk_F)^{N_\ell}}{\gamma_1^{N_\ell-1}} \right)^2  }
\eeq
Further, for a fixed displacement field $u_D$, the Berry curvature in a given valley is exactly opposite to that of the conduction band (in the particle-hole symmetric model), and is given by
\beq
\Omega_{-,\tau}(\k) = - \Omega_{+,\tau}(\k) =  \frac{ N_\ell^2 \tau^z}{2} \left(  \frac{v^{2N_\ell} k^{2(N_\ell-1)}}{\gamma_1^{2(N_\ell-1)}} \right) \frac{u_D}{\left[ u_D^2 + \left( \frac{(vk)^{N_\ell}}{\gamma_1^{N_\ell-1}} \right)^2 \right]^{3/2}} 
\label{eq:berry2}
\eeq
Let us consider an SVP phase in the hole-doped regime, where the electrons are polarized in valley $\tau$.
This implies that all the holes are polarized in valley $\bar{\tau} = - \tau$.
Since the valence band is topologically trivial and thus has zero net Berry curvature when integrated over the entire BZ, we may just compute the net magnetization for the unoccupied states, and add an overall minus sign to our result find the net magnetization of the occupied states:
\beq
M_{-,\tau} &=& -\frac{e \mu_v}{\hbar} \int_{k \leq k_F} \frac{d^2k}{(2\pi)^2} \, \Omega_{-,\bar{\tau}}(\k) = \frac{e \mu_v}{\hbar} \int_{k \leq k_F} \frac{d^2k}{(2\pi)^2} \, \Omega_{+,\bar{\tau}}(\k) = -\frac{e \mu_v}{\hbar} \int_{k \leq k_F} \frac{d^2k}{(2\pi)^2} \, \Omega_{+,\tau}(-\k) \nn 
& = & \frac{N_\ell e \tau^z u_D \mu_v}{4 \pi \hbar} \left(\frac{1}{|u_D|} - \frac{1}{|\mu_v|}\right) =  \frac{N_\ell e \tau^z}{4 \pi \hbar} [u_D + \mu_v \, \text{sgn}(u_D)]
\eeq
where we have used that $\mu_v < 0$ for hole-doping. 
We can combine both scenarios by determining electron or hole doping by the sign of the chemical potential $\mu$, and  therefore write the orbital magnetization as:
\beq
M_{\tau} =  \frac{N_\ell e \tau^z }{4 \pi \hbar} \left[u_D - |\mu| \, \text{sgn}(u_D) \right]
\eeq
Using Eq.~\eqref{eq:Mtog}, for a fully spin and valley polarized phase, this implies a valley g-factor of
\beq
g_v = \frac{N_\ell e}{2 \pi |n| \hbar \mu_B}\left[u_D - |\mu| \, \text{sgn}(u_D) \right]
\label{eq:gv}
\eeq
which is Eq.~(4) in the main text.

For $\lambda_I \neq 0$, $g_v$ can be obtained by simple replacement of $u_D \to u_D + \lambda_I s^z \tau^z$, given the spin and valley sectors.  
Since the flavor-polarized phases require large displacement fields, $|u_D| \gg \lambda_I$.
Therefore, in practice $g_v$ remains almost unaffected by SOC in the regime of interest. 

\subsection{Numerical evaluation}

\begin{figure}[!t]
    \centering
    \includegraphics[width = 0.55\textwidth]{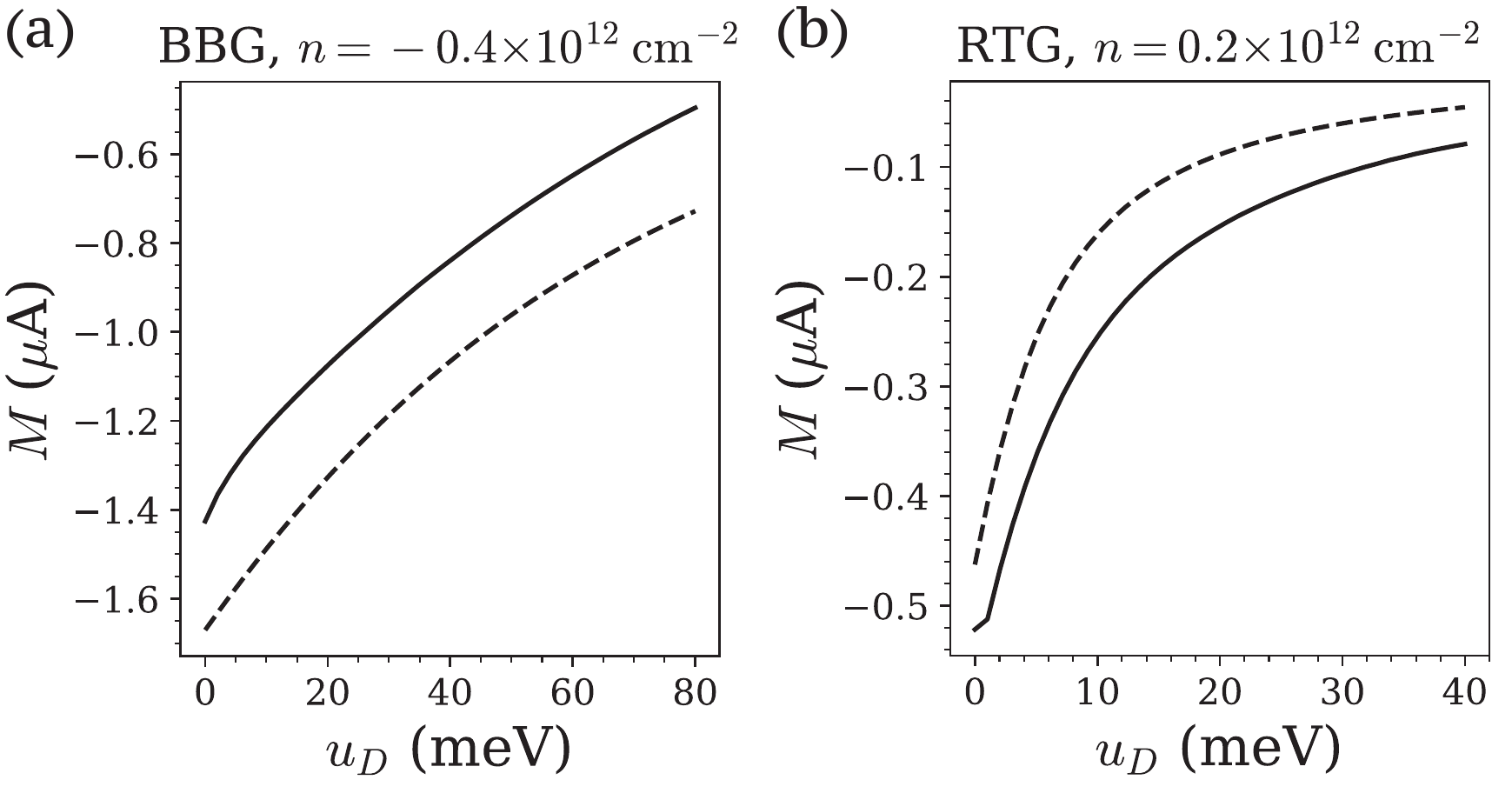}
     \caption{Orbital moment of BBG and RTG at two typical filling. We show the analytical result for the $2$-band model in Eq.~(4) and the numerical result obtained from Eq.~\eqref{eq:Msupp} for the $2N_{\ell}$-band model without SOC.}
    \label{fig:orbital_moment}    
\end{figure}

In this subsection, we elaborate on our numerical calculation of the orbital $g$-factor $g_v$, based on the orbital magnetization formula in Eq.~(4). 
We consider the non-interacting $2N_\ell$ band model in the absence of SOC. 
Taking $h_{4-band/6-band}$ as $h(\mathbf{k})$ and the corresponding Bloch wavefunction $\ket{u_{m,\mathbf{k}}}$ allows us to compute the magnetization of the active (partially filled) band in the $\tau$ valley at certain filling $n$, denoted as $M_0^{\tau}(n)$ such that $\sum_{\mathbf{k}} n_F(\varepsilon_m(\k)) = n$ (the superscript $0$ refers simultaneously to the non-interacting band-structure and no spin-orbit coupling). 
We note that $M_0^{K'}(n) = -M_0^K(n)$ as required by time-reversal symmetry.

In computing the magnetization density for symmetry-broken metals, we restrict ourselves to the parameter regime where the SVP/PFP phase is energetically favored, i.e. $E_{SVP/PFP} < E_{SVL}$.
As discussed earlier, the self-consistent Hartree Fock calculation gives us the one-electron covariance matrix $P_{\tau, \tau^{\prime}; s,s^\prime}^{m, m^{\prime}}(\mathbf{k})$ in the band basis.
We use this covariance matrix to compute the occupation of the active band $m$ in each valley
\beq
n_{\tau} = \sum_{\mathbf{k}} \Tr \left[ P_{\tau, \tau; s, s^\prime}^{m,m}(\mathbf{k}) \right]
\eeq
The total magnetization of the Hartree-Fock ground state can be approximated by 
\begin{equation}
    M[P_{\tau, \tau^{\prime}; s,s^\prime}^{m, m^{\prime}}(\mathbf{k})] = \sum_{\tau}M_0^{\tau}(n_{\tau})
\end{equation}
which can be easily converted to the valley $g$-factor shown in Fig.~3(b) using Eq.~\eqref{eq:Mtog}.

In Fig.~\ref{fig:orbital_moment}, we compare the analytical result in Eq.~(4) to the numerically computed value of $M_0^{\tau}(n_{\tau})$. 
The agreement between analytics and numerics is reasonably good, and the crucial conclusion that the magnetization decreases with increasing $u_D$ continues to hold for the microscopically accurate band structure.

Finally, we justify a couple of approximations used in this calculation.
First, we have neglected SOC in the evaluation of $M_0^{\tau}(n_{\tau})$, and therefore the valley $g$-factor.
This approximation is reasonable, as SOC does not alter the wavefunction and only contributes a constant energy shift to the active band in the doping range of interest, as discussed in the main text (also see Fig.~2). 
Further, we also neglected the redistribution of Berry curvature due to interaction. 
Though this effect might be important in some moiré systems \cite{SenthilChern}, in chiral-stacked graphene multilayers, interaction does not mix different bands at small doping.
Therefore, the interacting ground states are well-approximated as Slater determinants which are partially or fully flavor polarized.
For such states, it is reasonable to use the non-interacting wave-functions to calculate the Berry curvature.



%% file: main_PRL_final.bbl
\begin{thebibliography}{74}
\bibitem [{PHs()}]{PHsym}%
  \BibitemOpen
  \href@noop {} {}\bibinfo {note} {This statement is exact only in the limit of
  perfect particle-hole symmetry, but it is an excellent approximation for
  CSG.}\BibitemShut {Stop}%
\end{thebibliography}
